\documentclass[letterpaper]{sfchem}
\usepackage{graphicx}

\begin{document}

\title{Molecular Abundances in Barnard 68}

\author{James Di Francesco\inst{1,2} \and Michiel R. Hogerheijde\inst{1,3} \and 
William J. Welch\inst{1} \and Edwin A. Bergin\inst{4}} 
\institute{Radio Astronomy Laboratory, 601 Campbell Hall, University of California, Berkeley, Berkeley, CA, 94705-3411, U.S.A. \and National Research Council, Herzberg Institute of Astrophysics, 5071 West Saanich Rd., Victoria, BC V9E 2E7 Canada \and Steward Observatory, The University of Arizona, 933 N. Cherry Ave., Tucson, AZ, 85721-0065, U.S.A. \and Harvard-Smithsonian Center for Astrophysics, 60 Garden St. MS 42, Cambridge, MA, 02138-1516, U.S.A.} 
%Short author list here: Surnames only please (no initials)
\authorrunning{Di Francesco, Hogerheijde, Welch, Bergin}
%Short title here:
\titlerunning{Abundances of Molecular Species in Barnard 68}

\maketitle 

\begin{abstract}
Abundances for C$^{18}$O, CS, NH$_{3}$, H$_{2}$CO, C$_{3}$H$_{2}$, 
and N$_{2}$H$^{+}$ and upper limits for the abundances of $^{13}$CO 
and HCO$^{+}$ are derived for gas within the Bok globule B68 using 
our own BIMA array data, single-dish data from the literature, and 
Monte Carlo radiative transfer models.  B68 has had its density 
structure well determined, removing a major uncertainty from abundance 
determinations.  All abundances for B68 are lower than those derived 
for translucent and cold dense clouds, but perhaps only significantly 
for N$_{2}$H$^{+}$, NH$_{3}$, and C$_{3}$H$_{2}$.  CS depletion toward 
the extinction peak of B68 is hinted at by the large offset between 
the extinction peak and the position of maximum CS line brightness.  
C$^{18}$O and N$_{2}$H$^{+}$ abundances derived here are consistent 
with recently determined values at positions observed in common.

\keywords{ISM: molecules --- ISM: abundances --- ISM: globules}

\end{abstract}

\section{Introduction}

We examine the abundances of several molecular species in the 
Bok globule Barnard 68 (B68; Barnard 1919), located at 125~pc 
in Ophiuchus (de Geus, de Zeeuw, \& Lub 1989) with no evidence 
for internal star formation.  Figure 1 shows an $R$-band image 
of B68 obtained from the Aladin sky atlas, revealing its compact
morphology.  Using the extinction of the background Galactic bulge 
population, Alves, Lada, \& Lada (2001; ALL01) found a radial 
column density profile for B68 that was well matched with that 
of an isothermal Bonnor-Ebert sphere, specifically one with a 
near-critical center-to-edge density contrast of 16.5 (Bonnor 
1956; Ebert 1955).  This robust density determination removes a 
major uncertainty towards determining the molecular abundances in
B68.  Recently, Bergin et al.\/ (2002) used single-dish maps of 
B68 in conjunction with models incorporating the ALL01 Bonnor-Ebert 
density structure to ascertain radial variations in C$^{18}$O and 
N$_{2}$H$^{+}$ abundance, with the lowest values occurring toward 
the highest extinctions.  We explore further abundances of these 
and other species in B68, also incorporating the ALL01 density
structure into models.  

\begin{small}
\begin{figure}[ht]
\resizebox{\hsize}{!}{\includegraphics{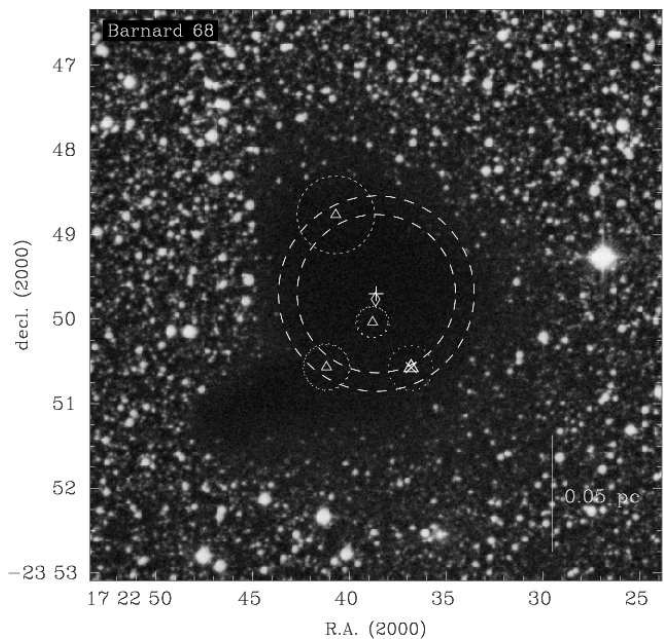}}
\caption{ESO MAMA $R$-band image of B68 from Aladin.  The extinction peak 
is denoted by a diamond.  The central pointing position of the BIMA data
is shown as a cross and the primary beam FWHMs at 89.2 GHz and 110.2 GHz 
are shown respectively as outer and inner dashed circles.  The positions 
and beam FWHMs of single-dish line observations from the literature are 
shown respectively as triangles and dotted circles.  The ``X" denotes the 
position of B68 from Clemens \& Barvainis (1988).  
\label{fig1}}
\end{figure}
\end{small}

\section{Observations}

Table 1 summarizes the data used to determine molecular abundances 
in B68, listing the lines used, the source observatory, the angular 
resolution, and the angular offset of the observation from the 
extinction peak.  Figure 1 shows the position observed using the 
BIMA millimeter interferometer, approximately 4\arcsec\ north of 
the extinction peak.  Figure 1 also shows the minimum and maximum 
FWHM sizes of the BIMA primary beams at the frequencies of $^{13}$CO 
1--0 and HCO$^{+}$ 1--0.  Line emission was not detected with the 
BIMA array at any observed frequency or position toward B68.  Table 
1 also lists studies where detections of lines were made toward B68 
from single-dish observations having angular resolutions $<$60\arcsec.  
Figure 1 shows the resolutions of these data and the positions where 
line characteristics were reported.  Note that these positions, 
typically the locations of peak line intensity, vary widely across 
B68. 

\section{Abundances} 

Abundances were estimated using the 1D Monte Carlo code of Hogerheijde 
\& van der Tak (2000) to calculate radiative transfer, 
molecular excitation, and line emission through a model of B68, and 
varying abundances until the output matched the observed data.  Our 
B68 model had the thermal and density structures from ALL01, i.e., an 
isothermal sphere of T = 16 K with an outer radius of 12 500 AU and a 
``Bonnor-Ebert parameter" $\xi_{\rm max}$ of 6.9 (i.e., center-to-edge 
density ratio of 16.5).  The central density is $\sim$2.5 $\times$ 
10$^{5}$ cm$^{-3}$.  Constant radial abundances were assumed.  Although 
abundance gradients within B68 are quite possible (see Bergin et al.), 
they should not produce large differences in the derived abundances, 
given the small sizes of the beam widths relative to the spatial extent 
of B68.  The velocity field of the gas was assumed to have a turbulent 
line width of 0.4 km s$^{-1}$ (see Wang et al.) with no bulk motions.

To compare models with interferometer data, ``visibility" datasets were 
produced from the model cubes using the baselines from the original data.  
These datasets were inverted, cleaned and restored in the same manner as 
the actual data, and velocity-integrated line intensities calculated.  By 
properly accounting for spatial filtering, we estimated BIMA would have 
recovered only $\sim$20\% of the flux emitted from an ALL01 Bonnor-Ebert 
sphere in B68.  To compare models with single-dish data, the model cubes 
were convolved with Gaussians of widths appropriately representing the 
resolutions of the respective observations.  

% Table 1
\begin{table}
\begin{center}
\caption{Summary of Line Observations of B68}
\renewcommand{\arraystretch}{1.2}
\begin{tabular}[h]{ccccc}
\hline
& & Beam & Offset \\
Line  & Site & (\arcsec\ ~$\times$~~\arcsec) & (\arcsec) & Ref. \\
\hline
$^{13}$CO 1--0                    &  BIMA    & 17.1 $\times$ 4.7 & 4.03 & 1 \\
C$^{18}$O 1--0                    &  BIMA    & 17.7 $\times$ 4.5 & 4.03 & 1 \\
C$^{18}$O 2--1                    &   CSO    &  30 $\times$ 30   & 59.3 & 2 \\
CS 2--1                           &  FCRAO   &  46 $\times$ 46   & 66.2 & 3 \\
HCO$^{+}$ 1--0                    &  BIMA    & 18.5 $\times$ 7.0 & 4.03 & 1 \\
N$_{2}$H$^{+}$ 1--0               &  BIMA    & 17.7 $\times$ 6.6 & 4.03 & 1 \\
N$_{2}$H$^{+}$ 1--0               & Haystack &  18 $\times$ 18   & 16.3 & 4 \\
NH$_{3}$ (1,1)                    &Effelsberg&  40 $\times$ 40   & 54.5 & 5 \\
H$_{2}$CO 3$_{12}$--2$_{11}$      &   CSO    &  30 $\times$ 30   & 59.3 & 2 \\
C$_{3}$H$_{2}$ 2$_{12}$--1$_{01}$ & Haystack &  18 $\times$ 18   & 16.3 & 4 \\
\hline 
\end{tabular}
\label{tab1}
\end{center}
{References --- (1) this work, (2) Wang et al.\/ 1995,
(3) Launhardt et al.\/ 1998, (4) Benson, Caselli \& Myers 1998, 
(5) Lemme et al.\/ 1996.}
\end{table}

Table 2 lists the fractional abundances of the sampled molecular species
at various positions toward B68.  The BIMA data yield the first upper limits 
to the abundances of HCO$^{+}$ toward the extinction peak of B68, as well as 
upper limits to the abundances of $^{13}$CO, C$^{18}$O and N$_{2}$H$^{+}$ at 
the same location.  In addition, single-dish data of B68 from the literature 
yield abundance values of H$_{2}$CO, CS, NH$_{3}$, and C$_{3}$H$_{2}$ toward 
B68 for the first time, but at positions offset from the extinction peak.  
Finally, other single-dish data of B68 from the literature yield new abundance 
values of C$^{18}$O and N$_{2}$H$^{+}$ at some of these latter locations.

Table 2 lists fractional abundances from the literature for the species 
studied here, in clouds similar in character to B68.  The first set, for 
``translucent clouds," are those listed by Turner (2000) for round Clemens 
\& Barvainis clouds with edge-to-center visual extinctions of 2.0.  The 
second set, denoted for ``cold dense clouds," are those compiled by Ohishi, 
Irvine, \& Kaifu (1992) toward TMC-1 or L134N.  Turner notes that these 
abundances should not be regarded as universal as they can vary by an order 
of magnitude both between different translucent clouds and within larger, 
dense clouds.  Every value of molecular abundance we derive for B68 is 
less than the lowest value derived for other clouds.  However, only the 
N$_{2}$H$^{+}$, NH$_{3}$, and C$_{3}$H$_{2}$ abundances differ by over an 
order of magnitude, and so only these species may be arguably depleted in 
B68, at least at the locations observed.  However, Figure 1 and Table 2 
show that less-discrepant abundances are found at positions relatively far 
from the extinction peak but more-discrepant abundances are found at positions 
closer to the extinction peak (except notably NH$_{3}$).  This pattern 
suggests C$_{3}$H$_{2}$ may be also depleted by some process related to 
extinction, e.g., the sublimation of gas-phase molecules onto grains, as 
suggested by Bergin et al.\/ for C$^{18}$O and N$_{2}$H$^{+}$.  This same 
idea may also explain how CS and H$_{2}$CO appear relatively undepleted in 
the outer, less-extincted radii of B68.  Moreover, the lack of bright emission 
in CS near the extinction peak hints that CS may be also depleted at high 
extinction in B68.  

% Table 2
\begin{table}
\begin{center}
\caption{Abundances of Various Molecular Clouds}
\renewcommand{\arraystretch}{1.2}
\begin{tabular}[h]{ccccc}
\hline
& & Translucent \\
Species & B68 & Clouds & TMC-1 & L134N \\
\hline
$^{13}$CO      & $<$1.3(-7)  & \null   & 8.9(-7) & 8.9(-7) \\
C$^{18}$O      & $<$1.3(-7)  & \null   & 1.6(-7) & 1.6(-7) \\
C$^{18}$O      &   3.0(-8)   & \null   & 1.6(-7) & 1.6(-7) \\
CS             &   4.0(-10)  & 1.1(-9) &  1(-8)  &  1(-9)  \\
HCO$^{+}$      & $<$1.4(-10) &  2(-9)  &  8(-9)  &  8(-9)  \\
N$_{2}$H$^{+}$ & $<$1.3(-10) &  1(-9)  &  5(-10) &  5(-10) \\
N$_{2}$H$^{+}$ &   2.0(-11)  &  1(-9)  &  5(-10) &  5(-10) \\
NH$_{3}$       &   7.0(-10)  & 2.1(-8) &  2(-8)  &  2(-7)  \\
H$_{2}$CO      &   4.0(-10)  & 6.3(-9) &  2(-8)  &  2(-8)  \\
C$_{3}$H$_{2}$ &   1.2(-11)  & 3.6(-8) &  3(-8)  &  2(-9)  \\
\hline
\end{tabular}
\end{center}
{Note --- Derived from $^{12}$CO abundance assuming $^{12}$C/$^{13}$C
= 90 or $^{16}$O/$^{18}$O = 500.}

{Note --- $a(-b)$ denotes $a \times 10^{-b}$}
\end{table}

For their abundance estimates for B68, Bergin et al. assumed a Zucconi et al. 
temperature profile.  They found the C$^{18}$O abundance in B68 rises from very 
low values at $A_{V}$~$<<$~1 to a peak of 1~$\times$~10$^{-7}$ at $A_{V}$~=~2, 
and decreases to 1~$\times$~10$^{-9}$ at $A_{V}$~$>$~20, a contrast of $\sim$100.  
Also, they found the N$_{2}$H$^{+}$ abundance rises from very low values at 
$A_{V}$~$<<$~1 to a peak of 6~$\times$~10$^{-11}$ at $A_{V}$~=~3, and decreases to 
3~$\times$~10$^{-11}$ at $A_{V}$ $>$~20, a contrast of $\sim$2.  Any evidence of 
N$_{2}$H$^{+}$ depletion is remarkable, given its oft-described utility as a 
non-depleting probe of dense core interiors (e.g., see Tafalla et al.\/ 2002.)  

Adopting the same temperature profile in our models, derived abundances are 
increased by factors of 2-3 from those listed in Table 2.  Our interferometer 
data of C$^{18}$O and N$_{2}$H$^{+}$ do not provide much additional support for 
the abundance models of Bergin et al., except that the non-detection of compact 
line emission with BIMA suggests the low abundances found by Bergin et al.\/ are 
not due to the beam dilution of small clumps of relatively abundant material.  Our 
derived upper limits for C$^{18}$O and N$_{2}$H$^{+}$ from the BIMA data are not 
particularly low, but they remain consistent with the still-lower values found 
by Bergin et al.\/ throughout the core.  

The single-dish C$^{18}$O and N$_{2}$H$^{+}$ data from the literature only 
pertain to one line-of-sight per transition toward B68, but provide strong 
support for the abundances derived by Bergin et al., again assuming the same 
Zucconi et al. temperature profile.  The C$^{18}$O literature data, from a position 
59\farcs4 offset from the extinction peak, yield an apparent abundance of 6--9 
$\times$ 10$^{-8}$, quite consistent with their 3--9 $\times$ 10$^{-8}$ abundances
at this position.  Also, the N$_{2}$H$^{+}$ literature data, from a position 
16\farcs3 offset from the extinction peak, yield an apparent value of 4--6 $\times$ 
10$^{-11}$, slightly larger than but still consistent with their 3--4 $\times$ 
10$^{-11}$ abundance at this position. 

\begin{acknowledgements}

JD was supported by the Radio Astronomy Laboratory.  MRH was supported by 
the Miller Institute for Basic Research in Science.  We thank Charles J. 
Lada, Jo\~ao Alves, Tracy Huard, and Jon Swift for valued discussions.
This research made use of the SIMBAD database and Aladin, both operated 
by CDS, Strasbourg, France.

\end{acknowledgements}

\end{document}